\documentclass[10pt]{article}
\usepackage[letterpaper]{geometry}
\usepackage{hicss}
\usepackage{times}
\usepackage[none]{hyphenat}
\usepackage{url}
\usepackage{latexsym}
\usepackage{minted}
\usepackage{indentfirst}
\usepackage{graphicx}
\graphicspath{{images/}}
\usepackage[
    style=apa,
  ]{biblatex}
\usepackage{float}

\usepackage{subcaption}
\usepackage[hidelinks]{hyperref}
\usepackage{booktabs}
\usepackage{tabularx}
\usepackage{listings}
\usepackage{array}

\newcolumntype{L}[1]{>{\raggedright\let\newline\\\arraybackslash\hspace{0pt}}m{#1}}
\newcolumntype{C}[1]{>{\centering\let\newline\\\arraybackslash\hspace{0pt}}m{#1}}
\newcolumntype{R}[1]{>{\raggedleft\let\newline\\\arraybackslash\hspace{0pt}}m{#1}}

\newcommand{\tabitem}{~~\llap{\textbullet}~~}

\title{Are Online Sports Fan Communities Becoming More Offensive?\\ A Quantitative Review of Topics, Trends, and Toxicity of r/PremierLeague}

\author{Muhammad Zeeshan Mazhar \\
University of Potsdam, \\
Germany \\
{\underline{ zeshmzm@gmail.com }} \\ \And
Tolga Buz \\
Hasso Plattner Institute \\
University of Potsdam, Germany \\
{\underline{ tolga.buz@hpi.de} } \\ \And
Yiran Su \\
Isenberg School of Management, \\
University of Massachusetts Amherst \\
{\underline{yiransu@isenberg.umass.edu}} \\ }

\date{}

\begin{document}
\maketitle
\begin{abstract}
Online communities for sports fans have surged in popularity, with Reddit's r/PremierLeague emerging as a focal point for fans of one of the globe's most celebrated sports leagues. This boom has helped the Premier League make significant inroads into the US market, increasing viewership and sparking greater interest in its matches. Despite the league's broad appeal, there's still a notable gap in understanding its online fan community. Therefore, we analyzed a substantial dataset of over 1.1 million comments posted from 2013--2022 on r/PremierLeague. Our study delves into the sentiment, topics, and toxicity of these discussions, tracking trends over time, aiming to map out the conversation landscape. The rapid expansion has brought more diverse discussions, but also a worrying rise in negative sentiment and toxicity. Additionally, the subreddit has become a venue for users to voice frustrations about broader societal issues like racism, the COVID-19 pandemic, and political tensions.
\end{abstract}

\subsubsection*{Keywords:}

Reddit, social media, social computing, hate speech, soccer

\section{Introduction}

The burgeoning online community of soccer fans has experienced exponential growth, particularly within forums dedicated to the Premier League such as Reddit's r/PremierLeague. 
As the most popular national soccer league and one of the most popular sports leagues globally, the Premier League commands a vast and enthusiastic following on social media \parencite{kemp2023most}. 
This online presence not only reflects the league's widespread appeal but also underscores the deep engagement of its international fan base.

The remarkable expansion of r/PremierLeague community deserves special attention. 
As a platform that allows for long-text information exchange, its growth is attributed to several factors: 
Reddit's increasing user base \parencite{weld2022makes}, the surging popularity of the Premier League on the platform, and the league's strategic push into the US market, where Reddit is predominantly used for a wide range of topics  \parencite{dean2023reddit}, including fan gathering and game information. 

In light of this impressive growth, %
there is much to learn about this community, the drivers behind its expansion, and its implications for Reddit, online sports fan communities, and the Premier League. 
Which topics does this community discuss and how can the interaction of users be described?
Are we able to identify trends that can potentially be generalized to the broader group of sports fans or even social media users in general?
Past studies have primarily focused on the content reflecting fans' identification with the team, often overlooking the intricate interplay between fans' emotions and platform utilization. \parencite{aswath2020analysing} 
In the wake of growing interest in online community well-being \parencite{rajadesingan2020quick, ghosh2021analyzing}, this gap suggests a need for a more detailed exploration of these dynamics.

Our work contributes to this area of research by investigating and quantifying the topics, sentiments, and trends posted across the r/PremierLeague subreddit. 
By doing so, it aims to illuminate the dynamics of online fan communities and offer valuable insights into their trends and health.

\section{Related Work}\label{sec:related-work}

The analysis of social media with different information extraction and Natural Language Processing (NLP) techniques has been conducted for a few years already, with studies yielding interesting insights about the reviewed datasets.
Sentiment analysis is a frequently used approach that has grown in popularity in the last years with the rapid increase in available online datasets, motivating a range of studies to apply this technique from different perspectives to better understand the dynamics within social media data \parencite{chakraborty2020survey}.
For example, a study has investigated the sentiment over time in Reddit discussions about vaccines during the Covid-19 pandemic \parencite{melton2021public}, and another study by \textcite{long2023gamestop} shows that this technique can also be employed in the context of financial discussions.

Another popular technique often applied in combination with sentiment analysis is topic modeling, which can be conducted using various algorithms such as Latent Dirichlet Allocation \parencite{blei2003latent} or using a pre-trained Large Language Model (LLM) such as BERTopic \parencite{grootendorst2022bertopic}.
This technique provides detailed insights about discussed themes rather than classifying a text as positive or negative, and can be used to analyse a new corpus exploratively \parencite{murakami2017corpus}.
A range of studies have employed topic modeling on social media data to identify relevant topics within discussions, e.g., in Covid-19-vaccine-related discussions \cite{chakraborty2020survey}, in tweets about e-learning during the pandemic-related lockdown \parencite{mujahid2021sentiment}, to identify contrasting opinions in discussions \parencite{stine2020comparative}, or to identify and extract sustainability practices from the discussions of a hiking community \parencite{Saaty2022Appalachian}.
Another piece evaluates Twitter discussions about autism and makes recommendations for developing meaningful public health agendas for autism based on the results \parencite{gabarron2023autistic}.

In the context of online sports (and e-sports) communities, these techniques have only been applied to a limited extent.
\textcite{wunderlich2022big} have measured changes in sentiment of soccer fans on Twitter over the duration of the match and extracted signals from these tweets to predict goals during matches, and \textcite{kasi2023unveiling} have evaluated how critically the Video Assistant Referee (VAR) technology was received among fans of the Premier League over time.
Another piece has reviewed Facebook data for a single Premier League season to identify differences between comments from male and female users \parencite{bagic2016sentiment}. 
Some research has also investigated how social media sentiment could be related to sports betting \parencite{hong2010wisdom, feddersen2017sentiment}.
\textcite{ning2022sports} have applied topic modelling on data from a Chinese sports Q\&A community.

Our work contributes to this area of research with a new perspective: 
we use multi-year data from r/PremierLeague to conduct an investigation of sentiment, topics, and toxicity in the community's discussions and review how this changes over the years. 
Specifically our analysis of topics within discussions identified as toxic yield interesting insights about Reddit's fan community for the world's largest sports league.

\section{Methodology}
\subsection{r/PremierLeague Dataset}
We obtain our raw dataset via the Pushshift API\footnote{Accessible via \url{https://api.pushshift.io}}. 
It consists of two separate files that contain either all submissions or all comments posted in the r/PremierLeague community from its creation in March 2011 until the end of 2022.
This yields a raw dataset of 70,028 submissions and 1,108,150 comments.
The dataset was retrieved in early 2023 before a change in Reddit's terms made the API practically inaccessible for open source solutions.

From the submission dataset, we use various features including the submission's \emph{id}, \emph{url}, \emph{title}, \emph{author}, time of creation (\emph{created\_utc}), number of comments (\emph{num\_comments}), \emph{score}, and \emph{link\_flair\_text} (i.e., a manually set category tag for each submission).
Regarding the comment dataset, features that we use include the comment's \emph{id}, \emph{parent\_id} (linking to the parent comment or submission), \emph{author} name, the upvote \emph{score}, \emph{created\_utc} (UTC timestamp), \emph{author\_flair\_text} (a tag chosen by the author, in this subreddit often the main team they are a fan of), and the comment's \emph{body} text.
This allows identifying most successful posts and comments, most active submission and comment authors, most popular category tags (link flairs), and activity over time.

There are various results from our exploratory analysis that may be of interest: 
As expected from a social media dataset, the scores for posts and comments are distributed exponentially

\subsection{Additional Datasets} 
To enrich our Reddit dataset and provide additional context, we utilize two other datasets:
Premier League game results \parencite{saife245_2023} and Twitter dataset \parencite{tirendazacademy_2022} about the 2022 FIFA World Cup labelled for sentiment.
The Premier League match results dataset contains results from 2017 to the end of 2022, covering 4,092 matches in total (1,579 wins, 1,579 losses, 934 draws). 
The dataset contains rich details for each match, including participating teams, date, time, venue, result, formations, referee, and sports betting odds.
The tweet dataset for the 2022 FIFA World Cup contains 22,000 tweets that have been labelled according to their sentiment. 
We use this dataset to measure how accurately our sentiment analysis approach works on datasets that are (1) retrieved from social media and (2) discussing topics around soccer.

\subsection{Analysis Techniques}

The main analysis techniques we use in this work are to some extent similar to related work presented in Section \ref{sec:related-work}, as we employ sentiment analysis and topic modeling.
We aim for a broader methodological coverage by including toxicity analysis and reviewing changes over time as the community and its US footprint grow.

\paragraph{Sentiment Analysis}
In line with related work, we apply sentiment analysis to identify whether the texts take a positive, neutral, or negative position.
In the first step, we clean the text data to remove the stop words and eliminate other phrases that are only specific to Reddit.
Additionally, special characters are stripped from the text and irrelevant words were removed by hand-made vocabulary created for our dataset (this mainly addresses certain texts that are automatically added to a large number of comments within the community such as links to provide feedback, or to read the FAQs or community guidelines). 

We utilize the BERTweet model \parencite{perez2021pysentimiento} for this task. However, before applying it to our dataset, the model is fine-tuned on the labelled 2022 FIFA World Cup Twitter dataset \parencite{tirendazacademy_2022}, which is pre-labelled for sentiment analysis, in order to optimize the model for our project's context.
The training parameters are a batch size of 32, a learning rate of 2e-5, and 3 epochs -- we then utilize the AdamW optimizer.
We use a train-validate-test dataset split of 80-10-10.
The fine-tuned model achieves sufficiently high F1-scores for the sentiment classification task on the test dataset to be considered accurate for our use case: 0.94, 0.90, and 0.95 (for positive, neutral, negative sentiment, respectively) and a weighted average F1-score of 0.93.

To test the robustness of the BERTWeet model for sentiment analysis, we manually label 100 randomly selected entries from the dataset, and compare it to the model's predictions. %
Over four runs (to test robustness), we measure a mean accuracy of approximately 82\%, which we consider sufficient for our use case.

\paragraph{Topic Modeling}
As explained above, we conduct topic modeling as another technique to provide a more granular overview of the most relevant topics that are being discussed in the community.
We use BERTopic \parencite{grootendorst_2022model} for this task - a model that uses BERT embeddings together with clustering algorithms to identify and group topics in text data. 
We apply data cleaning steps to prepare our dataset, including the removal of stop word and context-specific irrelevant words, and the replacement of special characters with whitespaces.
Furthermore, we generate subsets of our data for each year to better identify trends.
Our topic modeling pipeline then uses SBERT ("all-MiniLM-L6-v2") for embeddings, UMAP for dimensionality reduction, and HDBSCAN for clustering.
For SBERT we use default hyperparameters, for UMAP we set \emph{n\_neighbors} to 15, \emph{n\_components} to 5, \emph{min\_dist} to 0.0, and \emph{metric} to `cosine'.
Since our dataset has a different size for each year due to the increase in user activity, %
some parameters of the HDBSCAN algorithm are different for each year, i.e., we increase the parameters \emph{min\_cluster\_size} and \emph{min\_samples} gradually (we use default values for the remaining parameters). 

\paragraph{Toxicity Analysis}
Toxicity analysis is a technique less frequently used in related work, possibly due to it being relatively new approach enabled by deep-learning-based approaches.
In particular, the toxicity analysis approach that we use classifies a text as offensive, hateful, or abusive, if applicable.
This technique can be useful to understand the prevalence of toxic language within a community, and create a safer environment for online interactions. 
For our toxicity analysis, we apply a similarly rigorous preprocessing as for the previous tasks. This starts with data cleaning (stop word removal, replacement of special characters, and further filtering using a custom-built dictionary of irrelevant words and phrases.
We use a pre-trained RoBERTa model\footnote{Accessible via \href{https://huggingface.co/badmatr11x/distilroberta-base-offensive-hateful-speech-text-multiclassification}{\underline{this link to the Huggingface model hub}.}}. 
Our preliminary tests have identified this model as suitable for our use case, which a high correlation to human judgement in identifying toxic texts.
In an additional analysis, we conduct topic modelling on the texts classified as toxic. 
The results are shown in Section \ref{sec:toxic-topics}.

\subsection{Match Result Prediction}
Using the additional features generated by the techniques described above, we define a match result prediction task.
Our main motivation to do this is to use explainability metrics to better understand how the different features may contribute to a potential predictive value within the community's discussions.

For this purpose, we create weekly slices of the dataset so that each slice covers all posts made between two adjacent matches of the same team -- the motivation for this is to be able to analyze the subset of texts that are posted after or before a specific match.
As features, we select team name, opponent, sentiment labels, toxicity labels, and average score. 
We focus on the top six teams due to data availability.
These teams traditionally have the biggest fan base worldwide, which is reflected in our Reddit dataset. %

For the task, we train a random forest classifier to predict the outcome of a match (win, loss, draw) based on the discussions prior to the match. 
Hyperparameters are \emph{max\_depth}=30, \emph{min\_samples\_leaf}=2, \emph{min\_samples\_split}=2, \emph{n\_estimators}=205.

\section{Results}

\subsection{General Results}
\begin{figure} [ht]
    \centering
    \includegraphics[width=1\linewidth]{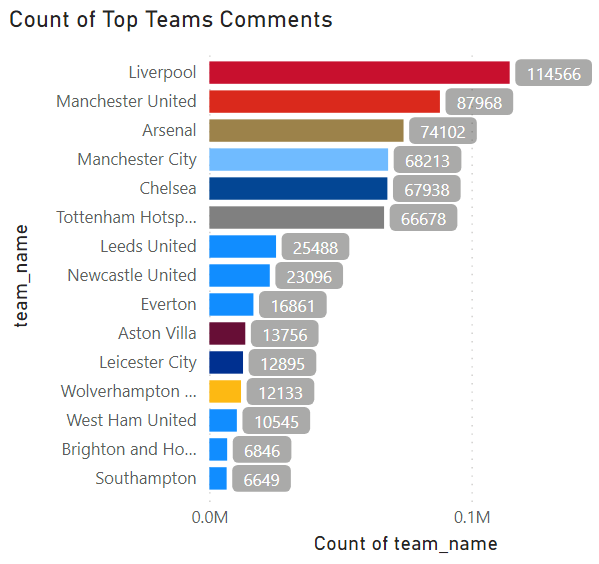}
    \caption{Most frequently chosen ``Author flairs'' as seen in comments data (users often use this flair to indicate their favorite team)}
    \label{fig:comment-flairs}
\end{figure}

As a first step, we conduct an exploratory analysis to better understand the feature distributions in our dataset.
An analysis of the ``author flairs'' (i.e., a tag that an author can choose for their profile, which is then shown in all comments they leave) in the comments dataset (Figure \ref{fig:comment-flairs}) shows the frequency of these flairs in the comments.
This can be seen as an indicator for how many fans each team has in the community, but is influenced by how actively each user writes comments.
The chart appears similar to an exponential distribution (which is often encountered in social media data), and shows that Liverpool is the most frequently encountered author flair (114,566), followed by Manchester United (87,968), Arsenal (74,102), Manchester City (68,213), Chelsea (67,938), and Tottenham Hotspur (66,678).
After these top six teams, there is a significant drop in frequency, indicating that this group, traditionally known as the top six of English soccer, is leading the Premier League regarding engagement and size of fan base.
We also reviewed all unique authors who have posted at least one comment to evaluate the frequency of each team-related flair:
Liverpool (3,484), Manchester United (3,241), Chelsea (2,971), Arsenal (2,913), Tottenham Hotspur (2,005), Manchester City (1,393), Leeds United (687), Newcastle	(675), Everton (650), West Ham (586).
These results indicate that fans of each team exhibit similar degrees of commenting activity.

\begin{figure} [ht]
    \centering
    \includegraphics[width=1\linewidth]{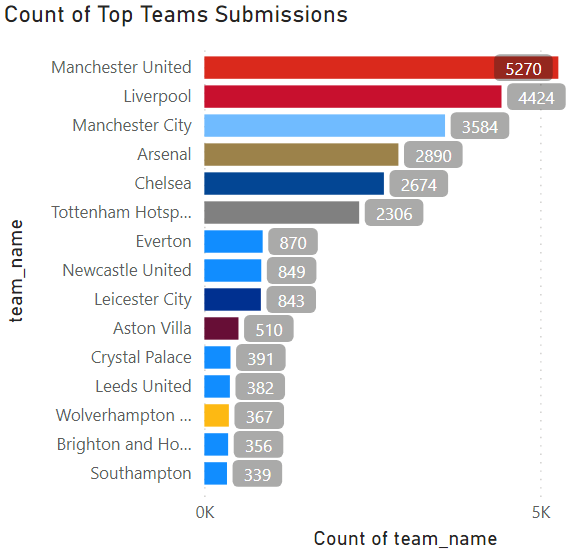}
    \caption{Most frequent submission flairs, indicating how often each team is posted about.}
    \label{fig:submission-flairs}
\end{figure}

Figure \ref{fig:submission-flairs} shows the distribution of the submission flairs.
In contract to the author flairs, this is the tag that an author has to choose specifically for a submission that they create to indicate the topic of the submission.
Again, the chart clearly shows the top six teams leading the Premier League, this time with Manchester United at the top, followed by Liverpool, Manchester City, Arsenal, Chelsea, and Tottenham Hotspur.
The fact that Manchester United and Liverpool are leading these charts and are mentioned together frequently in submissions and comments is in line with their long-standing rivalry, which is even documented on Wikipedia with a dedicated page (accessible via \href{https://en.wikipedia.org/wiki/Liverpool_F.C.%E2%80%93Manchester_United_F.C._rivalry}{this link})

In addition, our analysis shows a significant growth of the r/PremierLeague community with subscriber numbers doubling to tripling every year: 29,600 (end of 2017), 53,700 (end of 2018), 141,000 (end of 2019), 283,000 (end of 2020), 773,000 (end of 2021), 1.7 million (end of 2022).
At the time of writing in mid-2024, the subscriber count is at 3.8 million.
Naturally, with the growth of the user base, the amount of scoring, posting, and commenting activity increased as well.
This indicates a rapidly growing popularity of the subreddit and a likely adoption of the Premier League in the US, which is the main region of Reddit users (as explained above).

\subsection{Sentiment Analysis}
\begin{figure} [ht]
    \centering
    \includegraphics[width=1\linewidth]{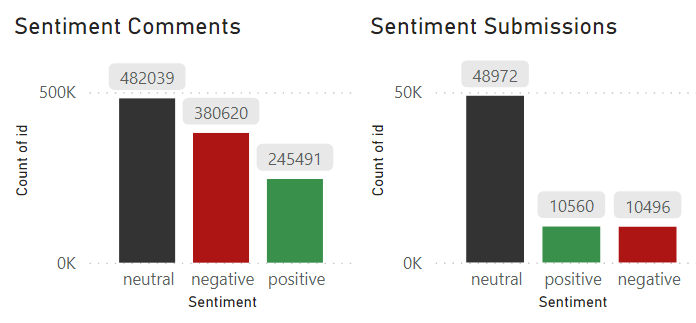}
    \caption{Distribution of sentiment classes across comments (left) and submissions (right)}
    \label{fig:sentiment-distribution}
\end{figure}

The sentiment analysis provides further insights about the dataset and r/PremierLeague's interactions. 
Figure \ref{fig:sentiment-distribution} show the distribution of the sentiment classes across comments and submissions.
It is clearly visible that for the submissions have a higher ratio of neutral texts than comments -- attribute this to the following reasons: (1) submissions are more strictly moderated than comments and (2) submissions are often links to external news articles and authors often copy the article headline into the submission title. 
While positive and negative sentiment are balanced in submissions (approx. 15\% each), comments show a stronger presence of negative (34\%) versus positive sentiment (22\%).
After a qualitative evaluation, we attribute this effect to the fact that whenever a specific team is discussed, there are often more fans of competing teams participating and voicing their criticism about the team rather than fans of the discussed team voicing positive opinions and support.

\begin{figure} [ht]
    \centering
    \includegraphics[width=1\linewidth]{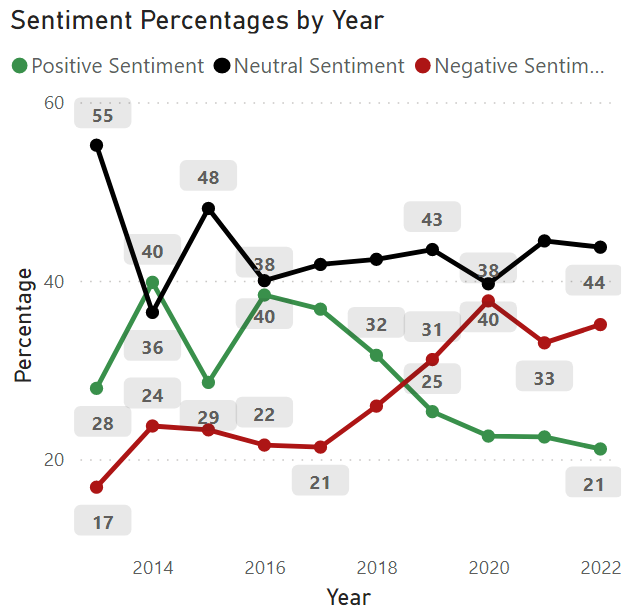}
    \caption{Percentage of Sentiment by Year}
    \label{fig:sentiment-trend}
\end{figure}

Figure \ref{fig:sentiment-trend} shows how the sentiment ratios changed over time.
Interestingly, the ratio of negative comments has increased significantly, from an average of 21\% for 2013--2017 to an average of 35\% for 2020--2022.
This could be caused by multiple effects, e.g., the rapid growth of the community combined with a weakening sense of community, the adoption of r/PremierLeague in the US, or changes in online social interaction due to the Covid-19 pandemic and lockdown.
This trend towards increasing negative sentiment is alarming and suggests that participating in the r/PremierLeague community has become less pleasant in recent years.

\subsection{Topic Modeling}
Our review of the identified topics throughout the years shows that most discussions focus on general soccer ``gossip'', while the top teams are subject to more criticism and negativity than the smaller teams.
We have clustered each year's most frequent topics using hierarchical clustering (HDBSCAN) and provide an manually-curated summary in Table \ref{tab:frequent-topics}.
Overall, the number of discussed topics has increased significantly over the reviewed years, which is in accordance with the growth in user numbers and activity.

\begin{table}[h]
    \centering
    \scalebox{0.8}{
    \begin{tabular}{L{0.12\linewidth}L{1\linewidth}}
    \toprule
        Year    &   Most Common Topics  \\
        \midrule
         2017   & 1. Team- \& match-specific discussions \\
                & 2. US broadcasters  \\
                & 3. Other leagues \& transfers \\
                & 4. Player-specific discussions (star players) \\
                & 5. Midfielders \& player signings  \\
                \midrule
         2018   & 1. Player-specific discussions (incl. injuries) \\
                & 2. Players in specific positions \& transfers \\
                & 3. Broadcasters \& streaming websites \\
                & 4. Team-specific fan groups  \\
                & 5. Video Assistant Refereee (VAR) \& referee decisions  \\
                \midrule         
         2019   & 1. Team- \& match-specific discussions (top teams) \\
                & 2. Match statistics \& offsides \\
                & 3. US-related topics (broadcasters, NFL, racism) \\
                & 4. Player-specific discussions \\
                & 5. Team-specific discussions (smaller teams)  \\
                \midrule
         2020   & 1. VAR \& offsides \\
                & 2. Player-specific discussions (star players)\\
                & 3. US sports broadcasters \\
                & 4. Team- \& match-specific discussions (top teams) \\
                & 5. Specific team managers  \\
                \midrule
         2021   & 1. VAR, referees, penalties \\
                & 2. Team- \& match-specific discussions (top teams) \\
                & 3. Specific team managers \\
                & 4. Player-specific discussions (star players) \\
                & 5. Off-topic discussions (racism, Covid-19, Americans)  \\
                \midrule
         2022   & 1. Top team performance \&  managers \\
                & 2. Off-topic discussions (jokes, racism, homosexuality) \\
                & 3. Offsides, VAR, referee decisions \\
                & 4. Specific team managers \\
                & 5. General discussion (soccer, injuries, goals)  \\
                \bottomrule
    \end{tabular}
    }
    \caption{Overview of top five most common topics per year in the dataset (sorted by frequency; modelling results further clustered manually)}
    \label{tab:frequent-topics}
\end{table}

We identify a few particular patterns to highlight in the following:
There was a high presence of US-based broadcasters such as NBCSN, Peacock, and ESPN, with users complaining about prices and discussing broadcast times.
The occurrence of words like IPTV and stream point towards discussions about alternatives to traditional subscription services.
Starting in mid-2019, when VAR (Video Assistant Referee) came to the Premier League -- this is reflected in frequent discussions of this topic, together with other controversial referee decisions (e.g., offsides, penalties).
Covid-19 was a major topic in 2020, as the entire soccer world suffered due to it as many matches were either cancelled or postponed due to the lockdown -- the Premier League stopped all matches in March 2020 and resumed again in July 2020 behind closed doors. 

In summary, we see a growing presence of US-related topics, which together with the user growth suggests an increase in adoption and popularity of the Premier League in the US.
With the community's growth, discussions have become more diverse (e.g., with discussions of tactics, managerial decisions, and player transfers) and off-topic discussions became more prevalent.
It should be noted that racism, often connected to US-related topics is a frequent theme in the discussions.

\subsection{Toxicity Analysis}\label{sec:toxic-topics}
\begin{figure} [ht]
    \centering
    \includegraphics[width=1\linewidth]{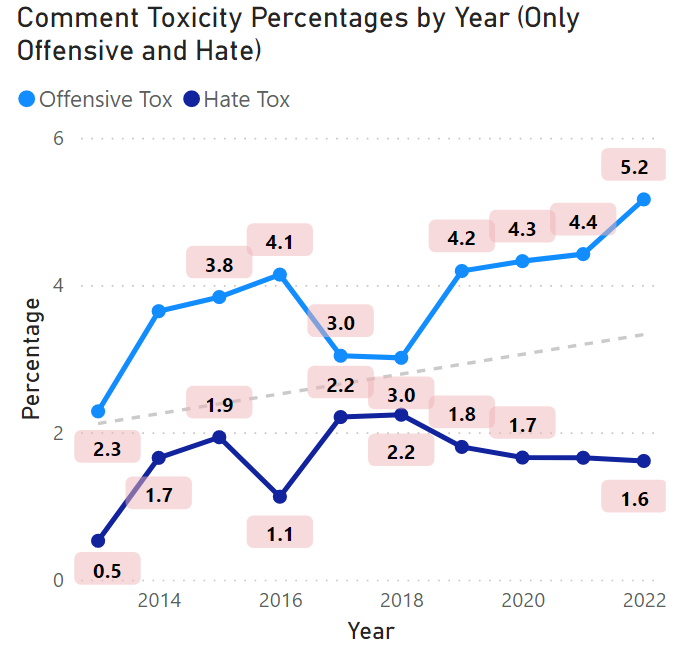}
    \caption{Ratio of texts classified as offensive or hateful per year by the RoBERTa toxicity model}
    \label{fig:toxicity-time}
\end{figure}

Our toxicity analysis results show that, on average, 4.5\% of all comments are labelled as offensive and 1.7\% as hate speech.
Again, we review the changes in these metrics over time, which is shown in Figure \ref{fig:toxicity-time}.
In accordance with the trends identified in the sentiment analysis, we can see that the amount of offensive texts has been increasing in the r/PremierLeague community constantly, from an average of 3.4\% (2013--2017) to 4.6\% (2020--2022).
This again indicates that the r/PremierLeague community has become more toxic and unpleasant over the last years, possibly caused by the effects explained above.

\begin{table}[h]
    \centering
    \scalebox{0.8}{
    \begin{tabular}{L{0.12\linewidth}L{1\linewidth}}
    \toprule
        Year    &   Toxic Topics  \\
        \midrule
         2017   & \tabitem Wayne Rooney \\
                & \tabitem Coutinho (transfer from Liverpool to Barcelona) \\
                & \tabitem Adam Johnson, Robin Van Persie (assault cases) \\
                & \tabitem Arsenal executive board \\
                & \tabitem Tottenham Hotspur vs. Arsenal  \\
                & \tabitem Liverpool fans (``scousers from Merseyside'') \\
                \midrule
         2018   & \tabitem FA and FIFA alleged bribery \\
                & \tabitem Marcos Alonso (injured Shane Long) \\
                & \tabitem Tottenham Hotspur vs. Arsenal  \\
                & \tabitem Australian people \\
                & \tabitem Manchester City \\
                & \tabitem Tottenham Hotspur fans \\
                \midrule
         2019   & \tabitem Racism \\
                & \tabitem Paul Pogba (poor performance) \\
                & \tabitem Liverpool vs. Manchester United  \\
                & \tabitem Tottenham Hotspur vs. Arsenal  \\
                & \tabitem Selhurst Park (stadium) \\
                & \tabitem British Colonial times and union jack \\
                & \tabitem Women's soccer equal pay discussion \\
                & \tabitem NBC Sports Network, USA \\
                & \tabitem Minors, rape, pedophilia, sexual assault \\
                & \tabitem Liverpool fans (``scousers from Merseyside'') \\
                \bottomrule
    \end{tabular}
    }
    \caption{Overview of most common topics to trigger toxic discussions per year for 2017 -- 2019 (sorted by frequency within each year)}
    \label{tab:toxic-topics-1}
\end{table}

We have applied topic modelling to the texts classified as toxic per year to identify the most controversial topics within the community -- \emph{which topics cause r/PremierLeague to become toxic}?
Tables \ref{tab:toxic-topics-1} and \ref{tab:toxic-topics-2} show the topics identified in our analysis to trigger toxic discussions for 2017--2019 and 2020--2022, respectively.
The results show an interesting mix of soccer-related and unrelated themes: 
Until 2019, there is a stronger focus on classic team rivalries (specifically, Tottenham Hotspur vs. Arsenal and Liverpool vs. Manchester United) and the behavior of selected players (e.g., Coutinho due to a transfer, Marcos Alonso injuring Shane Long through a foul).
In 2019, more general topics such as equal pay for Women's soccer, racism, British colonialism become prevalent. 
Also, US-related topics start occurring in 2019, e.g. the broadcaster NBC Sports Network, or a discussion of stereotypical American fans.
Overall, the number of different toxic topics increases over time and becomes broader due to an increase of more general topics such as racism or politics -- it should be noted that racism becomes a controversial topic every after starting in 2019.
The Video Assistant Refereee (VAR) becomes a frequent controversial topic after its introduction in 2020.
Furthermore, after 2020, the number of controversially discussed teams increases significantly, while the traditional rivalries seem not to be mentioned anymore.

\begin{table}[h!]
    \centering
    \scalebox{0.8}{
    \begin{tabular}{L{0.12\linewidth}L{1\linewidth}}
    \toprule
        Year    &   Toxic Topics  \\
        \midrule
         2020   & \tabitem Video Assistant Refereee (VAR) decisions \\
                & \tabitem Steven Gerrard, Frank Lampard, Wayne Rooney (midfielders) \\
                & \tabitem Black Lives Matter, racism and harassment in general and in soccer (e.g., players receiving death threats after missing penalties) \\
                & \tabitem US broadcasters (NBCSN, Peacock) \\
                & \tabitem The word ``negro'' and its link to Spanish language \\
                & \tabitem Fans of Leeds United, Liverpool, Arsenal, Tottenham Hotspur \\
                & \tabitem Politics (Communism, American government, Chinese government, Covid-19 cases) \\
                & \tabitem Australian and British people \\
                & \tabitem Racism incident with Luis Suarez and Patrice Evra \\
                \midrule
         2021   & \tabitem Racism against Black athletes in Europe \\
                & \tabitem VAR wrong decisions \\
                & \tabitem Arsenal, Liverpool, Newcastle, Manchester City, Chelsea, Manchester United, Everton, Tottenham Hotspur \\
                & \tabitem Ryan Giggs (personal life and cases) \\
                & \tabitem Stereotypical American fans supporting Premier League teams \\
                & \tabitem Complaints about vaccines \\
                & \tabitem Mikel Arteta \\
                \midrule
         2022   & \tabitem Racism, African and Black people \\
                & \tabitem VAR and controversial decisions \\
                & \tabitem Hillsborough tragedy, Liverpool \\
                & \tabitem Messi vs. Ronaldo \\
                & \tabitem Vladimir Putin and Saudi Arabia \\
                & \tabitem Leeds United, Arsenal, Liverpool, Everton, Tottenham Hotspur, Manchester United, Aston Villa, Chelsea \\
                & \tabitem The British monarchy \\
                & \tabitem Homosexuality in Britain and Premier League players \\
                & \tabitem Darwin Nunez, Jordan Pickford, Harry Kane, Ryan Giggs, Harry Maguire, and Gareth Southgate (England head coach) \\
                \bottomrule
    \end{tabular}
    }
    \caption{Overview of most common topics to trigger toxic discussions per year for 2017 -- 2019 (sorted by frequency within each year)}
    \label{tab:toxic-topics-2}
\end{table}

\subsection{Prediction Results}
Our match result prediction model achieves a training accuracy of 90.83\% and a test accuracy of 60.16\% - while this indicates overfitting, we deliberately choose to do this, as our priority lies in analyzing the model's important features learned from the training data, rather than producing accurate predictions on unseen data.

For predicting different match outcomes using the team flairs in the community's discussions, our model achieves the following F1 performance scores: 
Loss 0.27, Draw 0.08, Win 0.74.
The model works much better in predicting wins than losses or draws.
To better understand which features are most helpful for the model to make its predictions, we conduct a grid search of different feature combinations to train the model with, and find that the model works best when limited to \emph{total\_score}, \emph{neutral} and \emph{negative} sentiment, and number of texts labelled as \emph{offensive}.
Furthermore, we review the Gini Importance (also known as ``mean decrease impurity'' or ``feature importance''), for each feature in the final Random Forest model and find that the texts' upvote has the strongest impact (0.35), followed by neutral (0.26) and negative (0.25) sentiment, and offensiveness (0.14) (with poor match results correlating with toxicity).

This indicates that the upvote activity of r/PremierLeague may contain the highest predictive value for assessing whether a team will win a match, i.e., posts with higher upvote score are more likely to predict a match outcome.
Positive sentiments from fans, although they bring hope and support, don't significantly contribute to predicting match outcomes. This is because these sentiments often come with inherent biases and lack critical analysis, making them unreliable for accurate predictions. 
By understanding this limitation, we can approach sentiment analysis more effectively, leading to better and more realistic predictions based on how communities interact. 

\section{Conclusion}
Our work yields various insights about the reviewed community: (1) the discussed themes and topics of the community and how they change over time, (2) the growing relevance of the Premier League and r/PremierLeague subreddit for Reddit's user base, which is mainly US-based, and with it the US sports market; (3) the increase of negative sentiment and toxicity in the community's discussions over time. 
We apply and combine various analysis techniques to better understand the underlying data and identify interesting patterns.

Our topic modeling analysis revealed that discussions predominantly revolve around top Premier League teams -- Liverpool, Manchester United, Manchester City, Arsenal, Chelsea, and Tottenham Hotspur. 
Conversations mainly focus on match results, player performances, and rivalries, reflecting these teams' prominence and success in the league. 
Additionally, the significant volume of conversations about US broadcasters such as NBCSN, Peacock, Xfinity, and other US-related topics (as shown in our analysis) confirms that a significant portion of the subreddit's users are located in the US. 
Other frequently discussed themes include the VAR (Video Assistant Referee) system, controversies arising from referee decisions, and the impact of Covid-19 during 2020 and 2021.

Our toxicity analysis highlights the challenges associated with the community's growth, including an increase in offensive and off-topic discussions. 
Offensive remarks, particularly about other users and teams, tend to surge with the growth of the fan base. 
Specific targets of offensive comments include certain teams and fan groups such as Liverpool, as well as off-topic subjects such as political figures and entities like Putin, Saudi Arabia, and the British monarchy. 
Alarmingly, racism is a prevalent topic within these offensive discussions, indicating that it may be an issue within the fan community and potentially within the league itself.

While the growing popularity of the Premier League and its community has many positive aspects, our research also sheds light on the negative side effects, such as increasing negativity and toxicity. 
These effects significantly influence how users experience online communities and may reflect a broader trend towards higher toxicity and negativity in both online and real-world communities. 
It is dangerous to normalize toxicity in online communities, as individuals often dismiss positive communication in toxic environments \parencite{kavanagh2020towards, wijkstra2023help}. 
This normalization of negativity can lead to a cycle where toxicity becomes a standard part of their online experiences \parencite{poeller2023suspecting}. 
We encourage future research to continue investigating these trends, with the hope of reversing them to foster more positive and wholesome communities and social exchanges.

\printbibliography

\end{document}